\documentclass[conference,10pt]{IEEEtran}

\usepackage{framed}
\usepackage{algorithm}
\usepackage[noend]{algpseudocode}
\usepackage{bm}
\usepackage{stackrel}
\usepackage{subfigure}
\usepackage{epsf}
\usepackage{amsmath,amssymb}
\usepackage{graphicx}
\usepackage{color}
\usepackage{cite}
\usepackage{multirow,tabularx}
\usepackage{ifthen}
\usepackage{epstopdf}
\input{epstopdf.sty}
\usepackage{times}

\input{epsf.sty}

\makeatletter
\def\BState{\State\hskip-\ALG@thistlm}
\makeatother

\newcommand{\hide}[1]{\ifthenelse{\boolean{false}}{#1}{}}

\newtheorem{theorem}{{\bf Theorem}}

\newtheorem{lemma}{{\bf Lemma}}

\newcommand{\qed}{\nobreak \ifvmode \relax \else
      \ifdim\lastskip<1.5em \hskip-\lastskip
      \hskip1.5em plus0em minus0.5em \fi \nobreak
      \vrule height0.75em width0.5em depth0.25em\fi}

\newcommand{\beq}{\begin{equation}}
\newcommand{\eeq}{\end{equation}}
\newcommand{\barr}{\begin{array}}
\newcommand{\earr}{\end{array}}

\newcommand{\benum}{\begin{enumerate}}
\newcommand{\eenum}{\end{enumerate}}

\newcommand{\bit}{\begin{itemize}}
\newcommand{\eit}{\end{itemize}}

\newcommand{\bc}{\begin{center}}
\newcommand{\ec}{\end{center}}

\newcommand{\bdes}{\begin{description}}
\newcommand{\edes}{\end{description}}

\newcommand{\bfig}{\begin{figure}}
\newcommand{\efig}{\end{figure}}

\newcommand{\bemq}{\begin{quote} \begin{em}}
\newcommand{\eemq}{\end{em} \end{quote}}

\newcommand{\bmp}{\begin{minipage}}
\newcommand{\emp}{\end{minipage}}

\newcommand{\EX}[1]{\mathbb{E}\left[{#1}\right]} 

\newcommand{\bsp}{\begin{slide*}}
\newcommand{\esp}{\end{slide*}}
\newcommand{\bsl}{\begin{slide}}
\newcommand{\esl}{\end{slide}}

\newcommand{\blem}{\begin{lemma}}
\newcommand{\elem}{\end{lemma}}
\newcommand{\bthm}{\begin{theorem}}
\newcommand{\ethm}{\end{theorem}}

\newcommand{\AoI}{\text{AoI}}
\newcommand{\VarD}{\text{VarD}}

\newcommand{\pr}[1]{\mathbf{P}\left[ #1 \right]}

\IEEEoverridecommandlockouts

\begin{document}

\title{Age-Delay Tradeoffs in Single Server Systems}
\author{Rajat Talak and Eytan Modiano
\thanks{The authors are with the Laboratory for Information and Decision Systems (LIDS) at the Massachusetts Institute of Technology (MIT), Cambridge, MA. {\tt \{talak, modiano\}@mit.edu} }
}

\IEEEaftertitletext{\vspace{-0.6\baselineskip}}

\maketitle

\begin{abstract}
Information freshness and low latency communication is important to many emerging applications. While Age of Information (AoI) serves as a metric of information freshness, packet delay is a traditional metric of communication latency. We prove that there is a natural tradeoff between the AoI and packet delay. We consider a single server system, in which at most one update packet can be serviced at a time. The system designer controls the order in which the packets get serviced and the service time distribution, with a given service rate.
We analyze two tradeoff problems that minimize packet delay and the variance in packet delay, respectively, subject to an average age  constraint.
We prove a strong age-delay and age-delay variance tradeoff, wherein, as the average age approaches its minimum, the delay and its variance approach infinity.
We show that the service time distribution that minimizes average age, must necessarily have an unbounded-second moment.
\end{abstract}

\section{Introduction}
\label{sec:intro}
Information freshness and low latency communication is gaining increasing relevance in many communication systems~\cite{2018_LowLatencySurvey_Fischione}.
Age of information (AoI) is a newly proposed metric of information freshness~\cite{2012Infocom_KaulYates}, while packet delay is a traditional metric of latency in communication. AoI measures the time since the last received fresh update was generated at the source, and is therefore a destination centric metric. It only accounts for packets that deliver fresh updates to the destination. Packet delay, unlike AoI, is a packet centric metric that takes into account the delay incurred by each packet in the system. In this work, we show that there is a natural tradeoff between the two metrics of AoI and packet delay.

AoI was first studied for the first come first serve (FCFS) M/M/1, M/D/1, and D/M/1 queues in~\cite{2012Infocom_KaulYates}.
Since then, AoI has been analyzed for several queueing systems~\cite{2012Infocom_KaulYates, 2014ISIT_KamKomEp, 2014ISIT_CostaEp, 2016X_Najm, sun_lcfs_better, Inoue17_FCFS_AoIDist, 2018_Ulukus_GG11, 2018ISIT_Yates_AoI_ParallelLCFS, 2011SeCON_Kaul, 2016_MILCOM_Ep_AoI_Buffer_Deadline_Replace, 2016_ISIT_Ep_AoI_Deadlines, 2018_ISIT_Inoue_AoI_Deadline}, with the goal to minimize AoI. Two time average metrics of AoI, namely, peak and average age are generally considered.

The advantage of having parallel servers, towards improving AoI, was demonstrated in~\cite{2014ISIT_KamKomEp, 2014ISIT_CostaEp, 2018ISIT_Yates_AoI_ParallelLCFS}. 
Having smaller buffer sizes~\cite{2011SeCON_Kaul, 2016_MILCOM_Ep_AoI_Buffer_Deadline_Replace} or introducing packet deadlines~\cite{2016_MILCOM_Ep_AoI_Buffer_Deadline_Replace, 2016_ISIT_Ep_AoI_Deadlines, 2018_ISIT_Inoue_AoI_Deadline}, in which a packet deletes itself after its deadline expiration, are two other considered ways of improving AoI.
In~\cite{sun_lcfs_better}, the LCFS queue scheduling discipline, with preemptive service, is shown to be an age optimal, when the service times are exponentially distributed. AoI for the LCFS queue with Poisson arrivals and Gamma distributed service was analyzed in~\cite{2016X_Najm}.

In most of these works, optimal update generation and service rate is sought that minimizes age, and in several, the question of determining a good queue scheduling discipline for AoI minimization is considered with interest. More recently, determining optimal update generation and service time distribution, that minimizes AoI, has been studied in~\cite{talak18_determinacy}. In~\cite{talak18_determinacy, talak19_AoI_heavytail}, we showed that for the LCFS queue with preemptive service (LCFSp) and G/G/$\infty$ queue, a heavy tailed service time distribution minimizes AoI. It is noteworthy that such a heavy tailed service maximizes packet delay for the LCFSp queue and packet delay variance for the G/G/$\infty$ queue, respectively.

\begin{figure}
  \centering
  \includegraphics[width=0.85\linewidth]{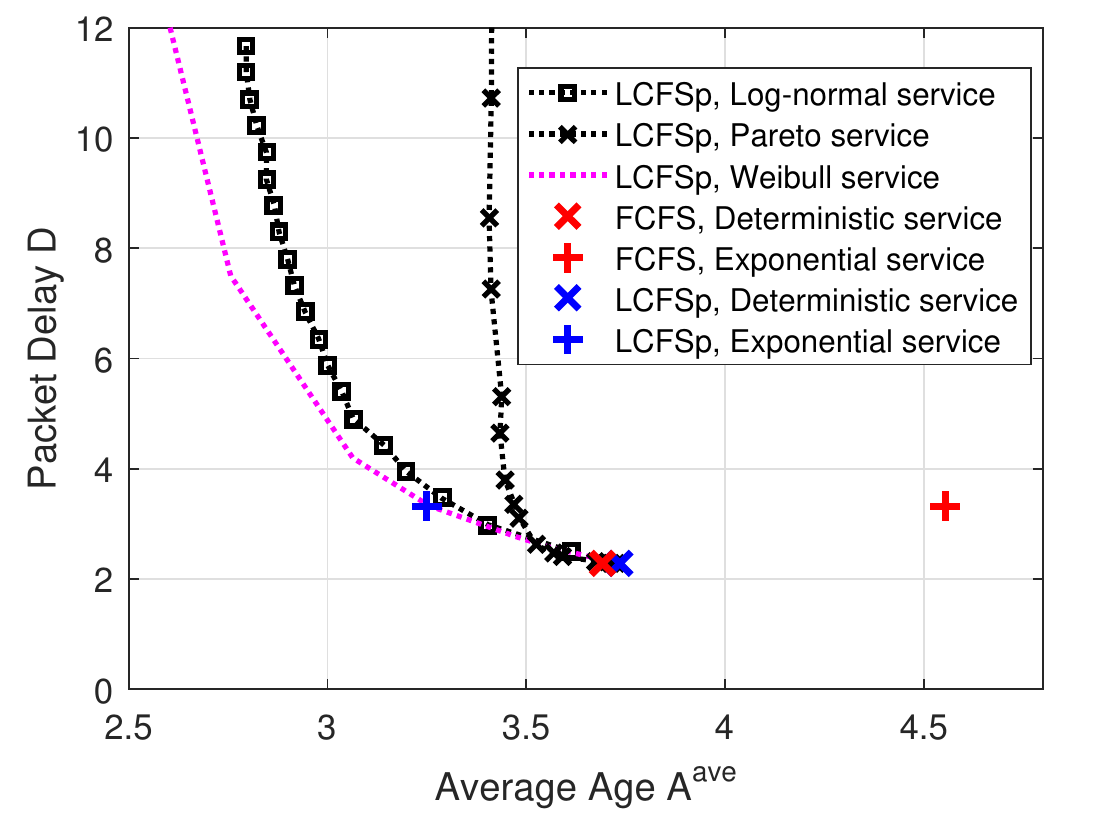}
  \caption{Plot of achieved age-delay points for various single server systems, Poisson packet generation at rate $\lambda = 0.5$, and service at rate $\mu = 0.8$. Scheduling disciplines: FCFS, LCFSp. Service time distributions: Deterministic, Exponential, and Heavy Tailed distributions in Table~\ref{tbl:heavy_tail}.}
  \label{fig:AoI_Delay_tradeoff1_plot1}
\end{figure}
\begin{table}
\caption{Heavy tailed service time distributions with mean $\EX{S} = 1/\mu$.}
\label{tbl:heavy_tail}
\begin{center}
\begin{tabular}{ |c|c|c| }
 \hline
 Name & Distribution & Free Parameter \\
 \hline
 Log-normal & $S = \exp\left(- \log \mu - \frac{\sigma^2}{2} + \sigma N\right)$ & $\sigma > 0$ \\
 & $N \sim \mathcal{N}(0, 1)$ & \\
 \hline
 Pareto & $F_S(x) = 1 - \left(\theta(\alpha)/x\right)^{\alpha}\mathbb{I}_{\{ x > \theta(\alpha) \}}$  & $\alpha > 1$ \\
 & $\theta(\alpha) = (\alpha - 1)/(\mu \alpha)$ & \\
 \hline
 Weibull & $\pr{S > x} = \exp\left\{- (x/\beta(k))^{k} \right\}$ & $k > 0$ \\
 & $\beta(k) = \left[ \mu \Gamma(1 + 1/k)\right]^{-1}$ & \\
 \hline
\end{tabular}
\end{center}
\end{table}
This points to a natural tradeoff between age and delay. In Figure~\ref{fig:AoI_Delay_tradeoff1_plot1}, we plot the achieved age-delay point under various queue scheduling disciplines and service time distributions. It appears that lower age can be achieved but only at a cost of higher delay. 
In this work, we prove that there is, in fact, a tradeoff between age and delay. 

We consider a single server system, that can service at most one packet at any given time. Generated updates are sent to this single server system. We assume that the system designer decides the queue scheduling discipline, i.e. the order in which the packets get serviced, and the service time distribution. Note that the service time distribution generally depends on the packet length, and therefore, a given packet length distribution can be induced on the generated update packets.

We consider the problem of minimizing packet delay, subject to an average age constraint, over the space of all queue scheduling disciplines and service time distributions, with a fixed mean service time budget of $1/\mu$.
For a given update generation process, we show that there is a strong age-delay tradeoff, namely, as the average age approaches its minimum, the delay approaches infinity. The same result holds also for packet delay variance, i.e. as the average age approaches its minimum, the variance in packet delay approaches infinity.
We also consider two restrictions on the system model, for which the age-delay tradeoff vanishes.

The system model and the age-delay tradeoff problems are described in Section~\ref{sec:model}. Strong age-delay and age-delay variance tradeoff is proved in~\ref{sec:age_delay_tradeoff}. In Section~\ref{sec:no_tradeoff}, we consider the two specific instances when the age-delay tradeoff vanishes, and conclude in Section~\ref{sec:conclusion}.


%

%

\section{System Model}
\label{sec:model}

A source generates update packets according to a renewal process, at a given rate $\lambda$. The update packets enter a queueing system. The server has rate $\mu$, and can service at most one update packet at any given time. The service times are independent and identically distributed across update packets.

The system designer has control over two things: It can decide the service time distribution, and it can decide the order in which the update packets get serviced. We assume that in determining the order of service, the scheduler is not privy to the service times of the individual packets.
The scheduler is also not allowed to drop any packets. Since we are concerned about age and packet delay, we will assume that the arrival rate is less than the service rate: $\lambda < \mu$.


%
We use $X$ to denote the inter-generation time of update packets with distribution $F_X$, and $S$ to denote the service time random variable, with distribution $F_S$. We use Minimize or $\min$, instead of the technically correct $\inf$, for ease of presentation. We now define the two latency metrics of average age of information and packet delay.

\subsection{Delay and Age of Information}
\label{sec:delay_aoi_def}
\begin{figure}
  \centering
  \includegraphics[width=0.9\linewidth]{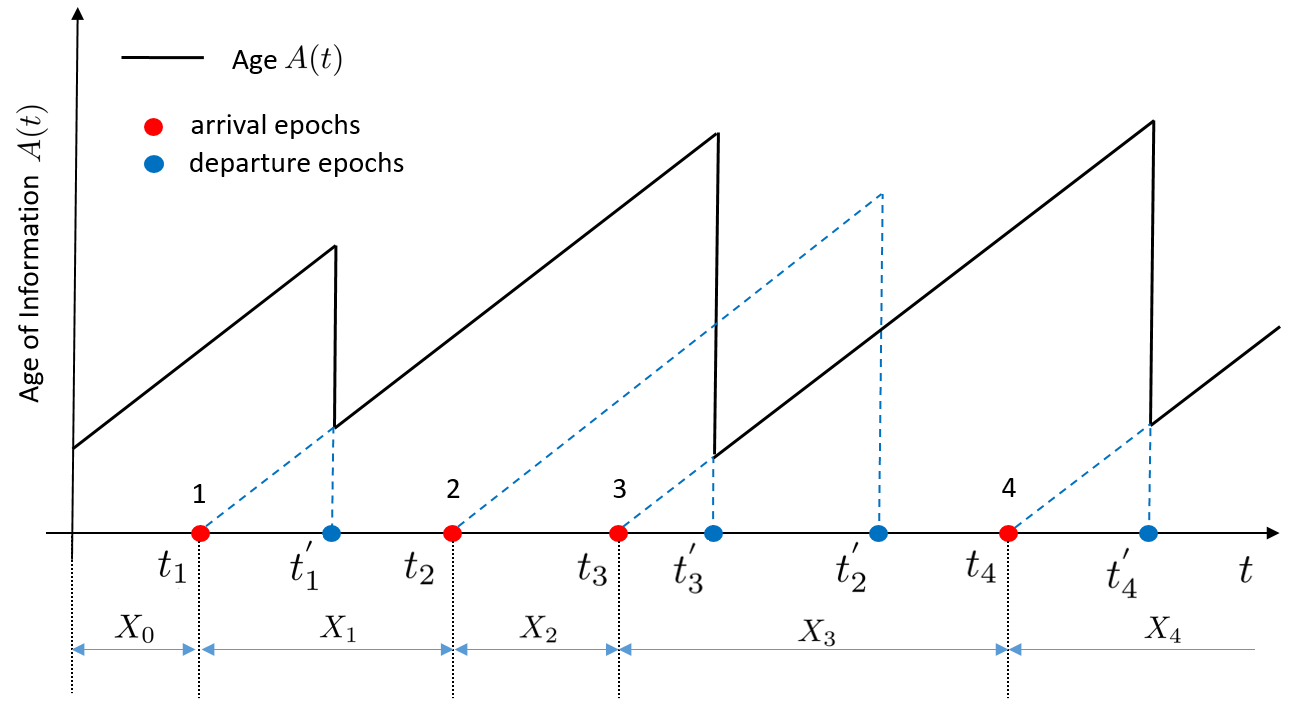}
  \caption{Age evolution in time. Here, $t_i$ and $t^{'}_{i}$ denote the generation and reception times of packet $i$.}
  \label{fig:age}
\end{figure}

Let the update packets be generated at times $t_1, t_2, \ldots$, and let the update packet $i$ reach the destination at time $t^{'}_{i}$. The update packets may not reach the destination in the same order as they were generated. In Figure~\ref{fig:age}, packet $3$ reaches the destination before packet $2$, i.e. $t^{'}_{3} < t^{'}_{2}$.
Delay for the $i$th packet is $D_i = t^{'}_{i} - t_{i}$, and the packet delay for the system is given by
\begin{equation}\label{eq:delay}
D = \limsup_{N \rightarrow \infty} \EX{\frac{1}{N}\sum_{i=1}^{N}D_i},
\end{equation}
where the expectation is over the update generation, service times, and scheduling discipline.
We skip a formal definition, but will use the notation $\VarD$ to denote variance in packet delay. For a single server queueing system, we note that $\VarD$ is lower-bounded by variance in service time distribution $S$.

Age of a packet $i$ is defined as the time since it was generated: $A^{i}(t) = (t-t_{i})\mathbb{I}_{\{ t > t_i\}}$,
which is $0$ by definition for time prior to its generation $t < t_i$.
Age of information at the destination node, at time $t$, is defined as the minimum age across all received packets up to time $t$:
\begin{equation}\label{eq:age_t}
A(t) = \min_{i \in \mathcal{P}(t)} A^{i}(t),
\end{equation}
where $\mathcal{P}(t) \subset \{1, 2, 3, \ldots \}$ denotes the set of packets received by the destination, up to time $t$.
Notice that AoI increases linearly, and drops only at the times of certain packet receptions: $t^{'}_{1}, t^{'}_{3}, t^{'}_{4}, \ldots$, but not $t^{'}_{2}$ in Figure~\ref{fig:age}. Such an age drop occurs only when an update packet with a lower age, than all packets received thus far, is received by the destination.
We refer to such packets, that result in age drops, as the  \emph{informative packets}~\cite{2014ISIT_KamKomEp}.
The average age is defined to be the time averaged area under the age curve:
\begin{equation}\label{eq:Aave}
A^{\text{ave}} = \limsup_{T \rightarrow \infty} \EX{\frac{1}{T}\int_{0}^{T}A(t) dt},
\end{equation}
where the expectation is over the packet generation and packet service processes.

We use the notation $D(F_S, \pi_Q)$, $\VarD(F_S, \pi_Q)$, and $A^{\text{ave}}(F_S, \pi_Q)$ to make explicit the dependency of delay, its variance, and average age, respectively, on the service time distribution $F_S$ and the queue scheduling policy $\pi_Q$.

\subsection{Age-Delay Tradeoff Problems}
\label{sec:tradeoff_prob_def}

We define two age-delay tradeoff problems. One, minimizes delay while the other minimizes delay variance, both over an average age constraint.
The age-delay tradeoff is defined as:
\begin{align}\label{eq:Aave_Delay_Tradeoff}
\begin{aligned}
T(\AoI) &= \underset{F_S, \pi_{Q}}{\text{Minimize}}
& & D(F_S, \pi_Q) \\
& \text{subject to} & & A^{\text{ave}}(F_S, \pi_Q) \leq \AoI, \\
& & & \EX{S} = 1/\mu.
\end{aligned}
\end{align}
Here, the function $T(\AoI)$ denotes the minimum delay that can be achieved for the single server queueing system, with an average age constraint of $A^{\text{ave}}(F_S, \pi_Q) \leq \AoI$. It might seem that both minimum age and delay could be attained simultaneously. We will show that, $T(\AoI) \rightarrow \infty$ as $\AoI$ approaches the minimum average age
\begin{align}\label{eq:Amin}
\begin{aligned}
A_{\min} &= \underset{F_S, \pi_{Q}}{\text{Minimize}}
& & A^{\text{ave}}(F_S, \pi_Q).
\end{aligned}
\end{align}
In~\cite{talak18_determinacy}, we proved such a result for the LCFS queues with preemptive service (LCFSp). In this work, we show that such a result holds, even when the system designer has an option of choosing the queue scheduling discipline $\pi_Q$. This does not follow trivially from the LCFSp result in~\cite{talak18_determinacy}, primarily because LCFSp is not known to be the optimal scheduling discipline for single server systems, especially when the service times are not exponentially distributed~\cite{BedewyISIT17_LIFO_opt, sun_lcfs_better}.
%

Variability in packet delay is also an important metric in system performance. We define the age-delay variance tradeoff problem to be:
\begin{align}\label{eq:Aave_DelayVar_Tradeoff}
\begin{aligned}
V(\AoI) &= \underset{F_S, \pi_{Q}}{\text{Minimize}}
& & \VarD(F_S, \pi_Q) \\
& \text{subject to} & & A^{\text{ave}}(F_S, \pi_Q) \leq \AoI, \\
& & & \EX{S} = 1/\mu.
\end{aligned}
\end{align}
Here, the function $V(\AoI)$ denotes the minimum delay variance that can be achieved for the single server queueing system, with an average age constraint of $A^{\text{ave}}(F_S, \pi_Q) \leq \AoI$. Counter to our intuition, we show that $V(\AoI) \rightarrow +\infty$ as \AoI~approaches its minimum value $A_{\min}$.

\section{Age-Delay Tradeoff}
\label{sec:age_delay_tradeoff}

Ideally, we would like to obtain every point on the tradeoff curves, i.e., completely characterize the functions: $T(\AoI)$ and $V(\AoI)$.
The following result motivates optimization of a linear combination of average age and packet delay, in order to achieve every point on the tradeoff curve.
\begin{framed}
\begin{theorem}
\label{thm:aoi_delay_min}
The points on the age-delay tradeoff curve $T(\AoI)$ can be obtained by solving
\begin{align}\label{eq:aoi_delay_min}
\begin{aligned}
& \underset{F_S, \pi_{Q}}{\text{Minimize}}
& & D(F_S, \pi_Q) + \nu A^{\text{ave}}(F_S, \pi_Q)\\
& \text{subject to} & & \EX{S} = 1/\mu,
\end{aligned}
\end{align}
for all $\nu \geq 0$. Similarly, the points on the age-delay variance tradeoff curve $V(\AoI)$ can be obtained by solving~\eqref{eq:aoi_delay_min}, by replacing $D(F_S, \pi_Q)$ with $\VarD(F_S, \pi_Q)$.
\end{theorem}
\end{framed}
\begin{IEEEproof}
The proof uses simple duality arguments, and is omitted due to space constraints.
\end{IEEEproof}
Theorem~\ref{thm:aoi_delay_min} motivates optimization of a latency metric that is a linear combination of average age and packet delay (or packet delay variance). This problem, however, is not easy to solve for the following reason: the delay is minimized with deterministic service times, while the variance in delay is minimized under FCFS service discipline~\cite{kingman_1962_var_min_queue}. The opposite holds for the average age: LCFSp queue scheduling policy with heavy tailed service distribution is known to achieve minimum age~\cite{talak18_determinacy, talak19_AoI_heavytail}. Thus, the delay term and the average age term in~\eqref{eq:aoi_delay_min} pull the decision variables in opposite directions.
%
In what follows, we prove that there is a strong age-delay tradeoff.

We say that a \emph{strong age-delay tradeoff} exists for $T(\AoI)$ if $T(\AoI) \rightarrow +\infty$ and $\AoI$ approaches $A_{\min}$. Conversely, \emph{no age-delay tradeoff} exists for  $T(\AoI)$ if the minimum average age and the minimum packet delay can be achieved simultaneously. Similar definition apply for age-delay variance tradeoff $V(\AoI)$.
\begin{figure}
  \centering
  \includegraphics[width=0.75\linewidth]{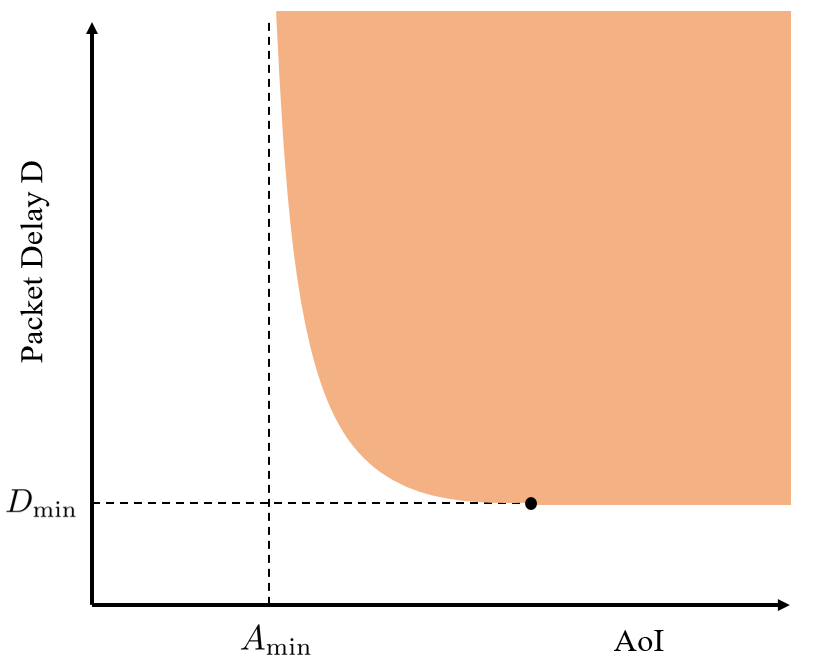}
  \caption{Illustration of strong age-delay tradeoff.}
  \label{fig:traeoff_illustration}
\end{figure}

Figure~\ref{fig:traeoff_illustration} illustrates a strong age-delay tradeoff. Note that this matches with our numerical results in Figure~\ref{fig:AoI_Delay_tradeoff1_plot1}.
In what follows, we prove a strong tradeoff between age-delay and age-delay variance.
We first derive the minimum average age $A_{\min}$, over the space of all scheduling policies and service time distributions.
\begin{framed}
\begin{lemma}
\label{lem:AoI_min}
The minimum average age $A_{\min} = \frac{1}{2}\frac{\EX{X^2}}{\EX{X}}$.
\end{lemma}
\end{framed}
\begin{IEEEproof}
The fact that $\frac{1}{2}\frac{\EX{X^2}}{\EX{X}}$ is a lower-bound on the average age, can be proved by pretending that each update packet spends zero time in the system, i.e. $t_i = t^{'}_i$. This provides a sample path lower bound for the age process. In this sample path, the age drops to $0$ at every $t_i$, and increases to $t_{i+1} - t_{i}$, just before dropping to $0$ again at $t_{i+1}$. The average age of this artificially constructed, lower-bound age process is $\frac{1}{2}\frac{\EX{X^2}}{\EX{X}}$, which implies $A_{\min} \geq \frac{1}{2}\frac{\EX{X^2}}{\EX{X}}$.

In~\cite{talak18_determinacy}, we proved that this age lower-bound is achieved by the LCFSp queue scheduling policy and heavy tailed service time distributions in Table~\ref{tbl:heavy_tail}: the Pareto, log-normal, and Weibull distributed service attain the lower-bound as $\alpha \downarrow 1$, $\sigma \uparrow +\infty$, and $k \downarrow 0$, respectively~\cite{talak18_determinacy}.
\end{IEEEproof}

%
We now prove the strong age-delay tradeoff and age-delay variance tradeoff.
\begin{framed}
\begin{theorem}
\label{thm:strong_age_delay}
There is a strong age-delay tradeoff and age-delay variance tradeoff, namely, $T(\AoI) \rightarrow +\infty$ and $V(\AoI) \rightarrow +\infty$ as $\AoI \rightarrow A_{\min}$.
\end{theorem}
\end{framed}
\begin{IEEEproof}
\textbf{1. Age-delay tradeoff:} First, note that the packet delay is given by $D(F_S, \pi_Q) = \frac{\lambda}{2}\frac{\EX{S^2}}{1-\rho} + \EX{S}$, where $\rho = \frac{\lambda}{\mu}$, for any scheduling policy $\pi_Q$ that does not use the individual packet service times to schedule it.
It, therefore, suffices to show that we have $\EX{S^2} \rightarrow +\infty$ as $\AoI \rightarrow A_{\min}$.

We first note that the average age $A^{\text{ave}}(F_S, \pi_Q)$, under any queue scheduling policy $\pi_Q$, is lower-bounded by the average age for the G/G/$\infty$ queue:
\begin{equation}\label{eq:lb}
A^{\text{ave}}(F_S, \pi_Q) \geq A^{\text{ave}}_{\text{G/G/}\infty}.
\end{equation}
This is because, in G/G/$\infty$ queue, an arriving packet is immediately put to service, and therefore, incurs no queueing delay. Due to this the average age for the G/G/$\infty$ queue serves as a lower-bound for any single server queue, in a stochastic ordering sense. Taking expected value yields~\eqref{eq:lb}.

From~\cite{talak18_determinacy}, we know the average age for the G/G/$\infty$ queue to be:
\begin{equation}\label{eq:Aave_gginf}
A^{\text{ave}}_{\text{G/G/}\infty} = \frac{1}{2}\frac{\EX{X^2}}{\EX{X}} + \EX{\min_{l \geq 0} \left( \sum_{k=1}^{l} X_k + S_{l+1} \right)}.
\end{equation}
Notice that the first term in~\eqref{eq:Aave_gginf} is nothing but $A_{\min}$. Therefore, as $\AoI \rightarrow A_{\min}$ in~\eqref{eq:Aave_Delay_Tradeoff}, it must be the case that $\EX{\min_{l \geq 0} \left( \sum_{k=1}^{l} X_k + S_{l+1} \right)} \rightarrow 0$. Lemmas~\ref{lem:gginf_implies_C} and~\ref{lem:C_implies_S2}, in Appendix~\ref{app:serv_dist_prop}, prove that $\EX{\min_{l \geq 0} \left( \sum_{k=1}^{l} X_k + S_{l+1} \right)} \rightarrow 0$ implies $\EX{S^2} \rightarrow +\infty$.

\textbf{2. Age-delay variance tradeoff:} Variance $\VarD(F_S, \pi_Q)$ is lower-bounded by the variance in service time $\EX{S^2} - \EX{S}^2$ for any queue scheduling policy $\pi_Q$. It therefore suffices to argue that as $\AoI \rightarrow A_{\min}$ we have $\EX{S^2} \rightarrow +\infty$, which we just proved to be true.
\end{IEEEproof}

In the proof, we essentially showed that $\EX{S^2} \rightarrow +\infty$ is a necessary condition for the average age to approach the minimum $A_{\min}$.
It seems counterintuitive at first that a strong tradeoff should exist between delay, or delay variance, and average age. However, a close examination reveals the following insight:

\emph{For age minimization it becomes necessary that the informative packets, the packets that reduce age, get serviced as soon as they arrive, while the non-informative packets, may incur as long a service time and queueing delay, as they do not matter in the age calculations. As we do this, the packet delay gets dominated by the delay of the non-informative packets, resulting in the two age-delay tradeoffs.}

We have assumed that the packet inter-generation time distribution to be fixed. 
The results in Lemma~\ref{lem:AoI_min} and Theorem~\ref{thm:strong_age_delay} imply that a strong age-delay tradeoffs will hold even if we could control the inter-generation time distribution $F_X$, with a mean budget of $\EX{X} = 1/\lambda$. The optimal $F_X$ would be deterministic as $A_{\min} = \frac{1}{2}\frac{\EX{X^2}}{\EX{X}^2} \geq \frac{1}{2}\EX{X}$, thus, making periodic generation of updates optimal. 

\section{Special Cases of No Tradeoff}
\label{sec:no_tradeoff}
In the previous section, we proved a strong age-delay and age-delay variance tradeoff. We now consider two scenarios of the single server system, for which the age-delay tradeoff vanishes, i.e. the minimum age and minimum delay can be attained simultaneously.

\subsubsection{Memoryless Service Times} Consider a system in which service times are exponentially distributed. The system designer has to decide only the queue scheduling policy $\pi_Q$ that solves~\eqref{eq:Aave_Delay_Tradeoff}.
We know from the works in~\cite{BedewyISIT17_LIFO_opt, sun_lcfs_better} that LCFSp minimizes average age, when the service times are exponentially distributed. The queueing delay $D(F_S, \pi_Q)$ remains the same, under any queue scheduling policy $\pi_Q$, that does not use individual packet service times to schedule~\cite{data_nets}. Thus, the minimum age and minimum delay is achieved simultaneously. A version of this result was proved in~\cite{2016_ISIT_YinSun_AoI_Thput_Delay_LCFS}.
\begin{framed}
\begin{theorem}[\cite{2016_ISIT_YinSun_AoI_Thput_Delay_LCFS}]
If service times are exponentially distributed, then there is no age-delay tradeoff. 
\end{theorem}
\end{framed}

\subsubsection{FCFS Queue Schedule} FCFS queue scheduling is used in many practical systems~\cite{talak18_Mobihoc, 2011SeCON_Kaul}. Periodic update generation is also known to reduces age in these systems. Consider the case of periodic update generation and FCFS queue scheduling. 

Deterministic service is known to minimize packet delay for the FCFS queue scheduling~\cite{data_nets}. It is also known that periodic generation and deterministic service minimized average age for the FCFS queue~\cite{Inoue17_FCFS_AoIDist, talak18_determinacy}: $A_{\text{D/D/1}}^{\text{ave}} \leq A_{\text{G/G/1}}^{\text{ave}}$. This gives us the following result.
\begin{framed}
\begin{theorem}
If the update generation is periodic and queue scheduling policy is FCFS, then there is no age-delay tradeoff.
\end{theorem}
\end{framed}

\section{Conclusion}
\label{sec:conclusion}

We considered a single server system, in which at most a single packet can be serviced at any given time. The system designer decides the order in which arriving packets get serviced and the service time distribution, with a given mean service time budget. We proved a strong age-delay and age-delay variance tradeoff, wherein as age approaches its minimum, the delay and its variance approach infinity.

We note the following reason for the tradeoff: For age optimality, informative packets, which reduce age, need to be services quickly, whereas a long service time and queuing delay can be incurred by other non-informative packets. As we do this, the packet delay, and its variance, get dominated by the delay of the non-informative packets. This leads to the age-delay tradeoff.

\bibliographystyle{ieeetr}

\begin{thebibliography}{10}

\bibitem{2018_LowLatencySurvey_Fischione}
X.~Jiang, H.~Shokri-Ghadikolaei, G.~Fodor, E.~Modiano, Z.~Pang, M.~Zorzi, and
  C.~Fischione, ``Low-latency networking: Where latency lurks and how to tame
  it,'' {\em Proc. of the IEEE}, pp.~1--27, Aug. 2018.

\bibitem{2012Infocom_KaulYates}
S.~Kaul, R.~Yates, and M.~Gruteser, ``Real-time status: How often should one
  update?,'' in {\em Proc. {INFOCOM}}, pp.~2731--2735, Mar. 2012.

\bibitem{2014ISIT_KamKomEp}
C.~Kam, S.~Kompella, and A.~Ephremides, ``Effect of message transmission
  diversity on status age,'' in {\em Proc. {ISIT}}, pp.~2411--2415, Jun. 2014.

\bibitem{2014ISIT_CostaEp}
M.~Costa, M.~Codreanu, and A.~Ephremides, ``Age of information with packet
  management,'' in {\em Proc. {ISIT}}, pp.~1583--1587, Jun. 2014.

\bibitem{2016X_Najm}
E.~{Najm} and R.~{Nasser}, ``Age of information: The gamma awakening,'' {\em
  ArXiv e-prints}, Apr. 2016.

\bibitem{sun_lcfs_better}
A.~M. Bedewy, Y.~Sun, and N.~B. Shroff, ``Minimizing the age of the information
  through queues,'' {\em arXiv e-prints arXiv:1709.04956}, Sep. 2017.

\bibitem{Inoue17_FCFS_AoIDist}
Y.~Inoue, H.~Masuyama, T.~Takine, and T.~Tanaka, ``The stationary distribution
  of the age of information in {FCFS} single-server queues,'' in {\em Proc.
  ISIT}, pp.~571--575, Jun. 2017.

\bibitem{2018_Ulukus_GG11}
A.~Soysal and S.~Ulukus, ``Age of information in {G/G/1/1} systems,'' {\em
  arXiv e-prints arXiv:1805.12586}, Jun. 2018.

\bibitem{2018ISIT_Yates_AoI_ParallelLCFS}
R.~D. Yates, ``Status updates through networks of parallel servers,'' in {\em
  Proc. {ISIT}}, pp.~2281--2285, Jun. 2018.

\bibitem{2011SeCON_Kaul}
S.~Kaul, M.~Gruteser, V.~Rai, and J.~Kenney, ``Minimizing age of information in
  vehicular networks,'' in {\em Proc. {SECON}}, pp.~350--358, Jun. 2011.

\bibitem{2016_MILCOM_Ep_AoI_Buffer_Deadline_Replace}
C.~Kam, S.~Kompella, G.~D. Nguyen, J.~E. Wieselthier, and A.~Ephremides,
  ``Controlling the age of information: Buffer size, deadline, and packet
  replacement,'' in {\em Proc. {MILCOM}}, pp.~301--306, Nov. 2016.

\bibitem{2016_ISIT_Ep_AoI_Deadlines}
C.~Kam, S.~Kompella, G.~D. Nguyen, J.~E. Wieselthier, and A.~Ephremides, ``Age
  of information with a packet deadline,'' in {\em Proc. {ISIT}},
  pp.~2564--2568, Jul. 2016.

\bibitem{2018_ISIT_Inoue_AoI_Deadline}
Y.~Inoue, ``Analysis of the age of information with packet deadline and
  infinite buffer capacity,'' in {\em 2018 IEEE International Symposium on
  Information Theory (ISIT)}, pp.~2639--2643, Jun. 2018.

\bibitem{talak18_determinacy}
R.~Talak, S.~Karaman, and E.~Modiano, ``Can determinacy minimize age of
  information?,'' {\em arXiv e-prints arXiv:1810.04371}, Oct. 2018.

\bibitem{talak19_AoI_heavytail}
R.~Talak, S.~Karaman, and E.~Modiano, ``When a heavy tailed service minimizes
  age of information,'' in {\em Submitted to {ISIT}}, Jul. 2019.

\bibitem{BedewyISIT17_LIFO_opt}
A.~M. Bedewy, Y.~Sun, and N.~B. Shroff, ``Age-optimal information updates in
  multihop networks,'' in {\em Proc. {ISIT}}, pp.~576--580, Jun. 2017.

\bibitem{kingman_1962_var_min_queue}
J.~F.~C. Kingman, ``The effect of queue discipline on waiting time variance,''
  {\em Mathematical Proceedings of the Cambridge Philosophical Society},
  vol.~58, no.~1, p.~163–164, 1962.

\bibitem{data_nets}
D.~P. Bertsekas and R.~G. Gallager, {\em Data Networks}.
\newblock Prentice Hall, 2~ed., 1992.

\bibitem{2016_ISIT_YinSun_AoI_Thput_Delay_LCFS}
A.~M. Bedewy, Y.~Sun, and N.~B. Shroff, ``Optimizing data freshness,
  throughput, and delay in multi-server information-update systems,'' in {\em
  Proc. {ISIT}}, pp.~2569--2573, Jul. 2016.

\bibitem{talak18_Mobihoc}
R.~Talak, S.~Karaman, and E.~Modiano, ``Optimizing information freshness in
  wireless networks under general interference constraints,'' in {\em Proc.
  {Mobihoc}}, Jun. 2018.

\end{thebibliography}

\appendix

\subsection{Properties of Service Time Random Variable $S$}
\label{app:serv_dist_prop}
Let $S$ be a continuous random variable with distribution $F_S$, with parameter $\eta$, such that $\EX{S} = 1/\mu$ for all $\eta$. We would like to derive conditions on $S$ such that
\begin{equation}\nonumber
\EX{\min_{l \geq 0} \left( \sum_{k=1}^{l} X_k + S_{l+1} \right)} \rightarrow 0,
\end{equation}
as $\eta$ approaches certain $\eta^{\ast}$,
for a given distribution $F_X$. We now derive an equivalent condition that only requires verifying certain properties of $F_S$.
\begin{framed}
\begin{lemma}
\label{lem:gginf_implies_C}
For $S_l$ and $X_k$ that are i.i.d. distributed according to $F_S$ and $F_X$, respectively, we have
\begin{equation}\label{eq:min_term}
\lim_{\eta \rightarrow \eta^{\ast}} \EX{\min_{l \geq 0} \left( \sum_{k=1}^{l} X_k + S_{l+1} \right)} = 0,
\end{equation}
if and only if, for all $x \geq 1/\lambda $, we have
\begin{equation}\label{eq:suff_cond}
\lim_{\eta \rightarrow \eta^{\ast}} \pr{S > x} = 0,~\text{and}~\lim_{\eta \rightarrow \eta^{\ast}} \EX{S \mathbb{I}_{\{ S < x\}}} = 0.
\end{equation}
\end{lemma}
\end{framed}
\begin{IEEEproof}
The fact that~\eqref{eq:suff_cond} imply~\eqref{eq:min_term} is proved in our recent work~\cite{talak18_determinacy, talak19_AoI_heavytail}. Here, we establish that~\eqref{eq:min_term} implies the conditions~\eqref{eq:suff_cond} on distribution $F_S$.

Let $Z = \min_{l \geq 0} \left( \sum_{k=1}^{l} X_k + S_{l+1} \right)$. We first lower-bound $Z$ as follow:
\begin{equation}\nonumber
Z = \min\{ S_1, X_1 + S_2, X_1 + X_2 + S_3, \ldots \} = \min\{ S_1, X_1 + Z' \},
\end{equation}
where $Z' = \min\{ S_2, X_2 + S_3, X_2 + X_3 + S_4, \ldots \}$. Since $Z' \geq 0$, we must have $Z \geq \min\{S_1, X_1\}$. Without loss of generality, we can loose the subscripts and write $Z \geq \min\{S, X\}$, where $S \sim F_S$ and $X \sim F_X$.

If $\EX{Z} \rightarrow 0$ as $\eta \rightarrow \eta^{\ast}$ then clearly $\EX{\min\{S, X\}} \rightarrow 0$ as $\eta \rightarrow \eta^{\ast}$. Now construct $\hat{X}$ such that:
\begin{equation}\nonumber
\hat{X} = \left\{ \begin{array}{cc}
                    0 &~\text{if}~X < 1/\lambda \\
                    1/\lambda &~\text{if}~X \geq 1/\lambda
                  \end{array}\right..
\end{equation}
Clearly, $\hat{X} \leq X$, and thus, $\min\{S, \hat{X}\} \leq \min\{S, X\}$, which implies $\EX{\min\{S, \hat{X}\}} \rightarrow 0$. Since $\hat{X}$ takes only two values, namely $0$ and $1/\lambda$, we have $\EX{\min\{S, \hat{X}\}} = \EX{\min\{S, 1/\lambda \}}\pr{X \geq \lambda}$. Now, $\pr{X \geq \lambda} > 0$ because $\EX{X} = 1/\lambda$. Further, $\pr{X \geq \lambda}$ does not depend on $S$, and therefore, is also independent of the parameter $\eta$. Therefore, $\EX{\min\{S, \hat{X}\}} \rightarrow 0$ implies $\EX{\min\{S, 1/\lambda \}} \rightarrow 0$. Using monotonicity of $\pr{S > x}$ in $x$ one can show that $\EX{\min\{S, 1/\lambda \}} \rightarrow 0$ implies
\begin{equation}\label{eq:nuclear2}
\lim_{\eta \rightarrow \eta^{\ast}} \EX{\min\{S, x\}} = 0,
\end{equation}
for all $x \geq 1/\lambda$.
Now, notice that
\begin{equation}\label{eq:nuclear3}
\EX{\min\{S, x\}} = \EX{S\mathbb{I}_{ \{ S < x \} }} + x \EX{\mathbb{I}_{ \{ S > x\} }},
\end{equation}
where we can ignore the equality case $S = x$ since $S$ is continuously distributed. Substituting~\eqref{eq:nuclear3} in~\eqref{eq:nuclear2} we obtain~\eqref{eq:suff_cond}.
\end{IEEEproof}

We now give a sufficient condition on the service time distributions $F_S$, parameterized by $\eta$, to have its second moment approach infinity.
\begin{framed}
\begin{lemma}
\label{lem:C_implies_S2}
For the parameterized, service time random variable $S$, we have $\lim_{\eta \rightarrow \eta^{\ast}} \EX{S^2} = +\infty$ if
\begin{equation}\label{eq:suff_cond2}
\lim_{\eta \rightarrow \eta^{\ast}} \pr{S > x} = 0,~\text{and}~\lim_{\eta \rightarrow \eta^{\ast}} \EX{S \mathbb{I}_{\{ S < x\}}} = 0,
\end{equation}
for all $x > x_0$, and some $x_0 > 0$.
\end{lemma}
\end{framed}
\begin{IEEEproof}
Let the two conditions~\eqref{eq:suff_cond2} hold for $S$. First, note that $\EX{S^2} \geq \EX{S^2 \mathbb{I}_{\{ S > x\}}} \geq x \EX{S \mathbb{I}_{\{ S > x\}}}$, for all $x > 0$. We can write $\EX{S \mathbb{I}_{\{ S > x\}}}$ as $\EX{S} - \EX{S \mathbb{I}_{\{ S < x\}}} \rightarrow 1/\mu$ as $\eta \rightarrow \eta^{\ast}$ by~\eqref{eq:suff_cond2} and the fact that $\EX{S} = 1/\mu$. Therefore, we have $\liminf_{\eta \rightarrow \eta^{\ast}} \EX{S^2} \geq x/\mu$ for all $x > x_0$. This can only be true if $\lim_{\eta \rightarrow \eta^{\ast}} \EX{S^2} = +\infty$.
\end{IEEEproof}


\end{document}